\documentstyle[preprint,aps]{revtex} 
\newcommand{\be}{\begin{eqnarray}}
\newcommand{\ee}{\end{eqnarray}}

\begin{document}
\title{Red Herrings and Rotten Fish} 

\author{
Geoffrey B. West$^{1,2}$
\footnote{To whom correspondence should be addressed. email: gbw@lanl.gov.}, 
Van M. Savage$^{1,2}$,
James Gillooly$^{3}$,  
Brian J. Enquist$^{4}$,
William. H. Woodruff$^{1}$,
James H. Brown$^{2,3}$
}

\address{
$^1$ 
Los Alamos National Laboratory,
Los Alamos, NM  87545, USA.\\
$^2$ The Santa Fe Institute,
1399 Hyde Park Road,
Santa Fe, NM 87501, USA.\\
$^3$ Department of Biology,
University of New Mexico,
Albuquerque, NM  87131, USA. \\
$^4$ Department of Ecology and Evolutionary Biology,
University of Arizona,
Tucson, AZ 85721, USA
 \\
}
\vspace{10.0cm}

\maketitle

\vspace{.5in}

\widetext

\pagebreak

A longstanding problem in biology has been the origin of pervasive quarter-power allometric
scaling laws that  relate many characteristics  of organisms  to body mass ($M$) 
across the entire spectrum of life  from  molecules and microbes to ecosystems
and mammals\cite{knut} - \cite{peters}.   In particular,  
whole-organism metabolic rate,  $B = aM^b$, where $a$ is a taxon-dependent
normalisation constant and $b\approx {3\over 4}$ for both  animals {\it and}
plants.  Recently Darveau {\it et al.}\cite{darveau}  (hereafter referred to as DSAH) 
proposed  a ``multiple-causes model"
for $B$ as ``the sum of multiple contributors to metabolism", $B_i$, which were assumed  to
scale as $B_i = a_i M^{b_i}$. They obtained for average values of $b$:
$0.78$ for the basal rate and $0.86$ for  the maximally active rate.  In this note we show that
DSAH contains serious technical, theoretical and conceptual errors,  including
misrepresentations of published data and of our work\cite{west1}-\cite{west3}.
We also show that, within experimental error, there is no 
empirical evidence for an increase in $b$ during aerobic activity as suggested by
DSAH. Moreover, since DSAH consider only  metabolic rates of mammals and make no attempt to
explain why metabolic rates for other taxa and many other attributes in diverse organisms also
scale  with quarter-powers (including most of their input data), their formulation is hardly the
``unifying principle" they claim. These  problems were not addressed in 
commentaries by Weibel\cite{weibel} and Burness\cite{burness}. 

{\it All} of the results of DSAH follow from their Eq. (2), $B =  a\sum c_i M^{b_i}$. Since, by
definition, the control coefficients\cite{control}, $c_i$, and exponents, $b_i$,  of the $i$th
process are dimensionless, this equation {\it must} be incorrect since it
violates the basic dimensional homogeneity constraint required of any
physical equation, namely, that all terms must have the same dimensions.  Their
Eq. (2) is therefore  meaningless. For example, it  necessarily gives different results when
using identical data but with different units for mass. To illustrate: we used their  data in
their Eq. (2) over the same mass range to obtain
$b$ for the basal rate; with mass  in Kg,  $b\approx 0.76$, when in g, $b \approx
0.78$, and when in pg,  $b\approx 1.08$.

DSAH merely state Eq. (2) without  proof, derivation or reference. 
General considerations  from standard control theory\cite{control}  expose some of the problems.
For a given metabolic state, consider
$B$ as  a function of the independent contributions, $B_i$: $B = B(B_i)$. By definition,
$c_i \equiv {\partial\ln B / \partial\ln B_i}$, leading to the
well-known sum rule, $\sum c_i = 1$, imposed by DSAH. Considering $B$ as
a function of mass, $B(M) = B[B_i(M)]$,  it  follows that 
$b = \sum c_i b_i$, where $b(M)\equiv d\ln B/d\ln M$, is the slope of the allometric  plot of
$\ln B$ vs. $\ln M$,  and $b_i(M)\equiv d\ln B_i/d\ln M$, that of $\ln B_i$
vs. $\ln M$. This is the formula that DSAH {\it should} have used to determine $b$. 
It is equivalent to the standard elasticity sum rule\cite{control}, $\sum c_i \epsilon_i = 0$,
with $\epsilon_i = b - b_i$. 
These equations are very general, requiring {\it no} assumptions about how $B$ and $B_i$
scale or whether the
$B_i$  are  in parallel or series.
In DSAH  contributions are added  in parallel so  $B = \sum  B_i$. Thus,
$c_i \equiv {\partial\ln B / \partial\ln B_i} = B_i/B = (a_i/a)M^{-\epsilon_i}$,  leading to 
$B = a\sum (c_i M^{\epsilon_i}) M^{b_i}$, which is the
correct, dimensionally homogeneous, version of Eq. (2). If $a$ and 
$a_i$ are constants, as assumed by DSAH, then, to be consistent, $c_i$ {\it must} scale as
$M^{-\epsilon_i}$; this $M$ dependence was omitted by DSAH.  Moreover, they  assume that
$c_i$ (and ${b_i}$) are also independent of $M$, so $b (= \sum c_i b_i$) must likewise not
depend on $M$, in  contradiction to their Eq. (2). This inconsistency is concealed in their
plots, which cover  {\it only} 3 orders of magnitude in $M$ over which $b$ is nearly
constant ($\sim 0.78$ for the basal case). However, when we extend their plots to the realistic
8 orders of magnitude spanned by mammals, the average  value of $b$ for the basal~rate
increases to
$\sim 0.85$ and, for the maximal rate, to $\sim 0.98$; both values are clearly inconsistent
with data\cite{knut}, \cite{peters}.

Even if DSAH had used correct  equations, there are many serious problems
with the  data and methodology used to estimate the $c_i$, $b_i$ and, in particular,
$b$ for maximal activity. Their treatment contains no statistical analysis: they  give no
confidence intervals for the  $c_i$ and  $b_i$, nor do they consider how, as data are combined,
the errors propagate to determine the confidence intervals for their estimate of $b$. Most 
$c_i$ quoted in DSAH are derived quantities. Almost none of the references cited actually
mention ``control coefficients" and DSAH's Methods section  gives insufficient information for
how they were derived.  For example,  $c_i$ should be obtained from {\it infinitesimal}  
responses in
$B$  rather than the large finite ones ($\sim 50\%$)   used by DSAH. Furthermore, the  ``data"
taken from Wagner\cite{wagner} for cardiac output,  alveolar ventilation and diffusion
capacities, are  {\it theoretical} estimates just for humans,  based only on a ``very simple
model"\cite{wagner} whose basic assumption is that ``$\dot V_{O_2}^{max}$ is ${O_2}$
supply-limited"\cite{wagner}, directly contradicting DSAH's central contention.  Several other 
$c_i$  are literally guesses (the $0.01$ values) and values for the $Ca^{++}$ pump
were obtained from  scaling of stride frequency of running mammals. In addition, a  factor of
0.8 is~arbitrarily introduced to rescale {\it some} 
$c_i$ to satisfy the sum rule,
$\sum c_i = 1$. The need for such a ``fudge" factor is hardly surprising given the empirical
uncertainties and theoretical misconceptions.

Also problematic is DSAH's contention that $b$ for   maximal activity 
is significantly larger than its basal value of
$\sim 0.75$. Both DSAH and Burness quote $b = 0.88\pm
0.02$ from Bishop\cite{bishop}, which is based on the {\it combined} data from only
$9$ mammals (including $2$ bats) and $6$ birds. Bishop obtained this value as the
exponent for $\dot V_{O_2}^{max}$ expressed as a power function of  {\it heart mass times
hemoglobin concentration} rather than body mass, $M$.  When expressed
as a function of $M$ his unadjusted data gives $b = 0.73\pm 0.04$ for the basal state and
$0.78\pm 0.08$ for
$\dot V_{O_2}^{max}$\cite{bishop}, in excellent agreement with previous
studies\cite{knut}, \cite{peters}. One of the most quoted of these  (though~ignored by DSAH)
is the  comprehensive study by Taylor {\it et al.}\cite{taylor} referred to by Weibel (who was a
co-author of the paper)\cite{weibel}. For 22 wild mammals (which the authors ``prefer  to
use....as the general allometric relationship for
$\dot V_{O_2}^{max}$"\cite{taylor} and which are of relevance here), they found $b = 0.79\pm
0.05$  and concluded that $\dot V_{O_2}^{max}$ ``is nearly a constant multiple of 10 times
resting metabolism $\dot V_{O_2}$, and scales approximately proportional to
$M^{0.75}$"\cite{taylor}. In his commentary Weibel cites this paper  as giving  $b = 0.86$ for
$\dot V_{O_2}^{max}$ but fails to remark that this  is derived  from  only 8 {\it domestic}
mammals and has very poor precision: the $95\%$ confidence limits, (0.611, 1.100), obviously
include $0.75$\cite{taylor}. 

Conservation of energy requires that summing  all ``ATP-utilising
processes"\cite{darveau} (linked in parallel) must give $B$: $B = \sum  B_i$. Consequently,
the  DSAH ``model" is  only a consistency check of  energy conservation, which must be
trivially  correct. As such, it cannot be in conflict with our theory. However,
in addition to the above problems, the processes that DSAH include in the sum  lead to  problems
of multiple-counting and thereby to a violation of  energy conservation. For example, they add
together contributions from cardiac output, alveolar ventilation, and pulmonary and
capillary-mitochondria diffusion as if they were independent and in parallel. But, as  shown in
the cartoon in Weibel's commentary\cite{weibel}, these processes  are, in fact, primarily in
series. The only reason DSAH obtain a  result for
$b$ in reasonable agreement with data   is that nearly all the exponents, $b_i$, have similar
values.
 
Since the  $b_i$ are simply taken from empirical data, DSAH's
formulation does not provide a fundamental  explanation for why $B$ scales non-linearly with
body mass. Such an
explanation would require models in which  the $b_i$ (and $c_i$) are derived from basic
principles. It is surely no accident that the $b_i$ cluster around ${3\over 4}$; understanding 
this  is the real challenge. Why, for example,  should molecular processes like  $Na^+$ and
$Ca^{++}$ pumps or  $ATP_{ase}$ activity  scale  other than linearly
with $M$? The simplest expectation, implicit in DSAH, would be that  the contributing
biomolecular processes, and therefore cellular metabolic rates, do not depend on body size, so
that $B$ would scale linearly with $M$. Moreover, nothing in DSAH~suggests  why cellular
level processes scale  differently {\it in vivo} than  {\it in vitro}. By contrast our
theory, based on scaling constraints inherent in  distribution networks and exchange surfaces,
correctly predicts that, when these constraints are lifted by removing cells from the
body and by culturing these cells for several generations,
{\it in vitro} cellular metabolic rates converge to a constant value - only {\it in vivo} do
they scale with body mass, as $ M^{-1/4}$\cite{west3}.

Finally, we respond to  DSAH's contention that our network theory is 
supply  rate-limiting and  cannot  accommodate metabolic scope. For a given metabolic state, 
scaling {\it  between} organisms of different sizes (varying $M$ with $a$ and $a_i$ fixed) is
indeed rate-limited by the network,  and this is the origin of quarter-powers.  However, {\it
within} an organism of a given size (fixed $M$), the absolute rate (as distinct from the
relative rate) of resource flow and power output (measured by the
$a$ and $a_i$, for example) is clearly {\it not} rate-limited by the network. Changes in supply
and demand cause the flow through the network to change accordingly,  as in any
transport system. A simplified analogy is the power output of automobile engines: this scales
with  size, but the power of any given vehicle can be varied  on demand by varying fuel
supply. Thus, our theory  accommodates metabolic scope in a natural and simple
way and could be extended to calculate the overall magnitude of increase.

To conclude: DSAH present their ``model" as an alternative to our theory for the origin of 
quarter-power  scaling  in biology.  Unlike their framework, however, our theory
offers a comprehensive, quantitative, integrated explanation for  allometric scaling, not just
of whole-organism metabolism in mammals, but also of many other characteristics in a wide
variety of animals, plants, and microbes.  It shows how the geometric, physical, and biological 
constraints on distribution networks and exchange surfaces with fractal-like designs give the
ubiquitous quarter-power scaling laws.  Our theory
explains why body size has such a powerful influence on biological structure and function at
all levels of organization, from the oxygen affinity of hemoglobin molecules and the density of
mitochondria, to the rate of fluid flow in the vessels of plants and animals, to ontogenetic and
population growth\cite{knut}-\cite{peters}.  

        By contrast, DSAH present a flawed model that purports to explain only how the
scaling of  whole-organism metabolic rate in mammals is related to the scaling of some of the
underlying processes at molecular, cellular, and organ-system levels.  Most importantly it
offers no explanation why any of these processes vary with body size, let alone why they should
exhibit their observed allometric exponents.  Thus, even if the errors were corrected, DSAH's
framework cannot  explain the quarter-power scalings of structures, rates, and
times that have so fascinated biologists for the last century.


\begin{thebibliography}{99}

\bibitem{knut}  Schmidt-Nielsen, K., {\it Scaling; Why is Animal Size So Important},  Cambridge
University Press, Cambridge, UK (1984).

\bibitem{calder} 
Calder III, W. A., Size, {\it Function and Life History}, Harvard Univ. Press, Cambridge, MA (1984).

\bibitem{peters}
Peters,  R. H. {\it The Ecological Implications of Body Size},
Cambridge Univ. Press, Cambridge (1983).

\bibitem{darveau} Darveau, C. A., Suarez, R. K., Andrews, R. D. and Hochachka, P. W.
  {\it Nature} {\bf 417}, 166 (2002).

\bibitem{west1} West,  G. B.,  Brown, J. H. and  Enquist, B. J.   {\it Science} {\bf 276}, 122 (1997).

\bibitem{enquist} Enquist, B. J., Brown, J. H. and West,  G. B. {\it Nature} {\bf 395}, 163 (1998)

\bibitem{gillooly1}  Gillooly, J. F., Brown, J. H., West, G. B., Savage, V. M.
and Charnov, E. L.  {\it Science}, {\bf 293}, 2248 (2001).

\bibitem{west3} West,  G. B.,  Brown, J. H. and Woodruff, W. H. {\it Proc. Nat. Acad. Sc.}
{\bf 99},  2473 (2002).

\bibitem{weibel} Weibel, E. R. {\it Nature} {\bf 417}, 166 (2002).

\bibitem{burness} Burness, G. P. {\it Science} {\bf 296},
1245 (2002).

\bibitem{control} Westerhoff, H. V. and van Dam, K. {\it Thermodynamics and Control of
Biological Free-Energy Transduction}, Elsevier, Amsterdam, The Netherlands (1987).

\bibitem{wagner} Wagner, P. D. {\it Respir. Physiol.} {\bf 93}, 221 (1993).

\bibitem{bishop} Bishop C. M. {\it Proc. R. Soc. Lond. B} {\bf 266}, 2275 (1999).

\bibitem{taylor} Taylor, C. R., Maloiy, G. M. O., Weibel, E. R., Langman, V. A., Kamau, J. M. Z.,
Seeherman, H. J. and Heglund, N. C. {\it Respir. Physiol.} {\bf 44}, 25 (1981).



\end{thebibliography}
\end{document}